# Graphene/Fluorinated Graphene Systems for a Wide Spectrum of Electronics Application

Antonova IV[1-3]*, Kotin IA[1], Kurkina II[3], Ivanov AI[1], Yakimchuk EA[1], Nebogatikova NA[1,2], Vdovin VI[1], Gutakovskii AK[1,2] and Soots RA[1]

[1]*Rzhanov Institute of Semiconductor Physics SB RAS, 630090, Novosibirsk, Russia*
[2]*Novosibirsk State Technical University, 630078, Novosibirsk, Russia*
[3]*Ammosov North-Eastern Federal University, 677000, Yakutsk, Russia*

**Abstract**

Heterostructures prepared from graphene and fluorographene (FG) using the technology of 2D printing on solid and flexible substrates were fabricated and studied. Excellent stability of printed graphene layers and, to a lesser degree, composite graphene:PEDOT:PSS layers were shown. Extraordinary properties of FG as an insulating layer for graphene-based heterostructures at fluorination degree above 30% were demonstrated. It is shown that the leakage current in thin (20-40 nm) films is normally smaller than $10^{-8}$ A/cm$^2$, the breakdown field being greater than $10^8$ V/cm. In hybrid structures with printed FG layers in which graphene was transferred onto, or capsulated with, an FG layer, an increase in charge-carrier mobility and material conductivity amounting to 5-6 times was observed. The spectrum of future applications of FG layers can be further extended due to the possibility of obtaining, from weakly fluorinated graphene (<20%), functional layers exhibiting a negative differential resistance behavior and, at fluorination degrees of 20-23%, field-effect-transistor channels with current modulation reaching several orders. Composite or bilayer films based on fluorographene and $V_2O_5$ or polyvinyl alcohol exhibit a stable resistive switching behavior. On the whole, graphene/FG heterostructures enjoy huge potential for their use in a wide spectrum of application, including flexible electronics.

## Introduction

Flexible electronics is presently considered as one of the main routes in the development of future electronic technologies. That is why a huge number of studies aimed at the search for materials with required mechanical properties and at the development of novel approaches and technologies for fabricating flexible electronic components are under way. Graphene possesses extraordinary mechanical properties: it can be extended by ~25% with a minimal change of many properties of this material (within 20%), including its electrical conductivity, and withstand bending and rolling influences [1]. Graphene oxide and, to a lesser degree, fluoro-graphene possess similar properties [2]. As the technologies for fabricating graphene and its derivatives further develop, a real breakthrough in the field of developing flexible electronic components is observed due to the unique combination of good mechanical properties of graphene and its high electrical conductivity, thermal conductivity, etc.

Development of methods for producing graphene derivatives with a wide spectrum of functional properties and, also, some other 2D materials provides a possibility for fabrication of a wide spectrum of device structures for flexible electronics and, hence, considerably widens the area of application of such materials and devices [3-5]. Printed technologies offer a cheap approach, alternative to lithography-based technologies, allowing the fabrication of device structures for flexible electronic components. It should be noted here that flexible electronics rapidly progresses not only due to the use of new materials but, also, due to the various combinations of novel and traditional materials and technologies. For instance, recently it has been shown possible to fabricate a non-volatile flexible memory based on a matrix with metal nanoparticles used as floating gates for small silicon flash-memory elements [6]. The area of application of flexible electronics includes touch-sensitive screens, various sensors and probes, including those for biological applications, radio-frequency identifiers and labels, photovoltaic elements, light-emitting diodes and electronic fabrics, super-capacitors, thin-film accumulators, and many other devices [7,8].

Presently, printing techniques are recognized as the most promising technologies for fabrication of flexible, expandable, bendable and rollable electronic components [3-8]. Initially, printing technologies were introduced as processes involving organic materials. However, practice has shown that organic materials in fact feature comparatively short service lives. Nonetheless, such materials have found applications due to the low production cost of their suspensions and device structures printed from those suspensions. Much interest in, and considerable practical demand for, the development of 2D printing processes is currently related with the breakthrough in the latter field, which has emerged due to the use of new inorganic materials such as graphene and its derivatives.

In the present study, we analyze the possibilities for fabricating lateral and vertical heterostructures from graphene and fluorographene (fluorination degrees 10 to 50%), the mutual influence of nearby layers on the electrical properties of main channel, and discuss possible applications of graphene- and fluorographene-based heterostructures. As the main materials, in the present paper we consider graphene and fluorographene (FG) suspensions, which allow their use in 2D printing technologies for fabrication of heterostructures. For initial graphene suspensions, a characteristic important for their applications is the flake thickness, and for fluorographene, the fluorination degree; the latter parameter alters the properties of FG material from semiconductors to insulators. Two-three-layered films and heterostructures were fabricated for our studies. Potential in using fluorographene as dielectric (gate dielectric and insulating layers) and functional films (negative differential resistance (NDR) structures, field-effect-transistor (FET) channels with current modulation reaching several orders, and composite graphene/FG layers with resistive switching behavior) is demonstrated.

## Experimental Studies

### The main process steps in the fabrication of grapheme-and fluoro-graphene-based structures

Recently, we have developed a graphene fluorination process which permits the preparation of materials with unique electrical (from conductors to insulators, depending on the fluorination degree) and optical properties [9-11]. The method is based on processing graphene or few layer graphene in an aqueous solution of hydrofluoric acid. We have showed that the resultant fluoro-graphene had a fluorination degree ranging from small values to ~25% ($C_4F$) (at this value, the transition into insulating state occurs) and to ~50% ($C_2F$). It was shown that the fluorination process could be used for passivation of flake edges and defects, for preparation of graphene-based tunnel-transparent multi-barrier systems, and for creation of FG-matrix-hosted quantum-dot arrays (at fluorination degree about 25%). However, in the fabrication of device structures, due to the high fluorographene stability (the binding energy of the C-F bond is 124 kcal/mol), difficulties with nanostructuring have emerged. The high binding energy of the C-F bond explains both the high thermal stability of the material (defluorination point ~450°C) and its high stability against various dissolving agents and harsh chemicals. It is well known that, for local defluorination of fluorographene, electron bombardment can be used [12]. On the other hand, as our experiments have shown, traditional lithographic methods fail to enable local removal of FG films at their open areal parts. That is why the strategy implying the use of fluorinated graphene suspensions for preparation of FG films and for implementation of 2D printing processes seems to be highly attractive. The possibility of fluorination of graphene suspensions was demonstrated in ref. [10]. Below, we will consider the main process steps in fabricating device structures from graphene/FG systems using fluorinated graphene suspensions and 2D printing processes. In some cases graphene oxide (GO) suspension prepared by modified Hummers method was used.

### Characterization Methods

A Solver PRO NT-MDT scanning microscope was used for obtaining atomic force microscopy (AFM) images of suspension flakes and film surfaces, and for determining the film thicknesses. The measurements were conducted in both contact and semi-contact mode. Raman spectra were recorded at room temperature, the excitation wavelength being 514.5 nm (2.41 eV argon ion laser). In order to avoid the heating of samples with laser radiation, the laser beam power was decreased to 2-3 mW. Scanning electron microscopy (SEM) images were obtained using a JEOL JSM-7800F scanning electron microscope with the energy of primary electrons equal to 2 keV. The film structure features were studied by means of high resolution transmission electron microscopy (HRTEM) using JEOL-4000EX and JEOL-2200 FS microscopes. For the HRTEM study, the graphene suspension was applied dropwise to thin (10 nm) carbon films and then dried for several hours. The sheet resistance of obtained films was studied using the four-probe JANDEL equipment and HM21 Test Unit. Capacitance-voltage (C-V) and current-voltage (I-V) characteristics of fabricated structures were measured using an E7-20 immitancemeter and a Keithley picoamperemeter (model 6485). Measured X-ray photoelectron spectroscopy (XPS) spectra permitted the study of the chemical composition of obtained films.

The films and the structures were fabricated by the 2D printing method. A Dimatix FUJIFILM DMP-2831 printer equipped with a DMC-11610 printing head with 16 nozzle carriers of about 20-μm diameter was used for printing. The printing process was implemented on both solid and flexible substrates. The solid substrates were silicon and oxidized silicon ($SiO_2$/Si) wafers with 300-nm oxide thickness covered with an APTES (3-Aminopropyl) triethoxysilane) film for ensuring good adhesion of water-based ink. As flexible substrates, Poly-ethylene-terephthalate (PET) substrates with an adhesive coating (Lamond), Epson-paper substrates for jet printing, and polyamide films (Kapton) were used. During the printing process, the substrate temperature was maintained at 60 °C. In some cases, hybrid structures in which printed dielectric layers were used for transfer, onto those layers, of graphene or few layer graphene flakes mechanically exfoliated from natural graphite or CVD-grown graphene or few layer graphene were employed.

## Results and Discussion

### Preparation of graphene suspensions

An important step in the preparation of fluorinated graphene suspensions is the creation of graphene suspensions. We used the electrochemical exfoliation of highly-oriented pyrolitic graphite (HOPG) with additional processing of exfoliated graphene flakes in a laboratory disperser. As a result of electrochemical exfoliation, a non-oxidized graphene flakes with typical lateral sizes of order one micrometer and typical thicknesses of 3-10 nm were obtained. After the additional processing and the filtration on track membranes, the lateral sizes of the flakes were within 400 nm, the latter value being 50 times smaller than the diameter of the printer nozzle. The thickness of graphene flakes could be further reduced to 1-3 monolayers, or 0.5-3 nm, depending on the subsequent purposes. For creation of transistor structures, in which case the mobility of charge carriers in the printed channels presents an important parameter, suspensions with 1-3-monolayer graphene flakes are necessary [13]. In other cases, in which only a high conduction of the graphene film is required, flake thicknesses of 0.5 to 3 nm are sufficient. A graphene suspension intended for preparation of ink, graphene particles contained in this suspension, a graphene film printed from this suspension, and a Raman spectrum taken from the printed film are shown in Figure 1.

Such an important property of the printed layers as their time stability deserves mention. Figure 2 shows the resistance of a printed graphene strip versus the duration of specimen storage in air without any additional protecting coating. For comparison, Figure 2 shows the resistances of similar strips printed out from a composite ink prepared from graphene and PEDOT:PSS (Poly(3,4-ethylenedioxythiophene):poly (styrenesulfonic acid)). The use of an additive of PEDOT:PSS, the most conducting polymer [14,15], solves a number of problems related with solution stability, nozzle clogging, and printed-layer drying. Stability of the properties of printed graphene:PEDOT:PSS structures deteriorated in comparison with the stability of printed graphene layers; yet, it substantially improved in comparison with the stability of PEDOT:PSS [14]. The printed graphene layers preserved their initial conductivity value over a period of one layer. Additional annealing of graphene layers at 200°C in argon ambient led to some reduction of their resistance, which value subsequently showed no substantial variations.

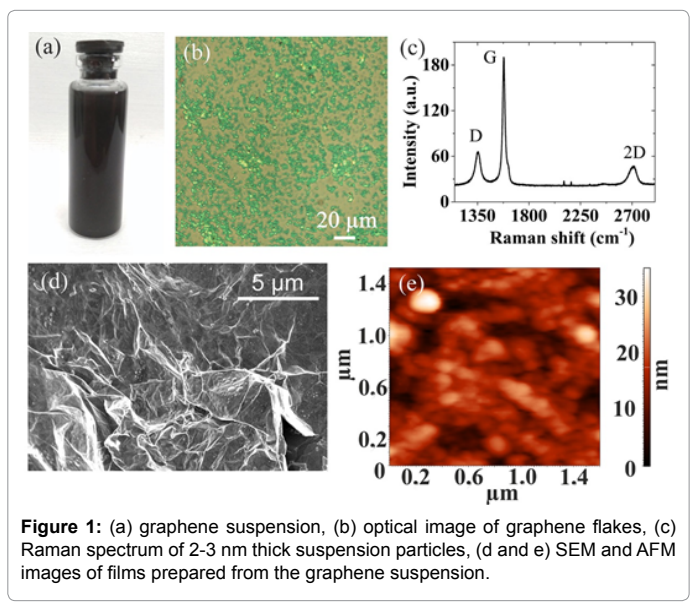

**Figure 1:** (a) graphene suspension, (b) optical image of graphene flakes, (c) Raman spectrum of 2-3 nm thick suspension particles, (d and e) SEM and AFM images of films prepared from the graphene suspension.

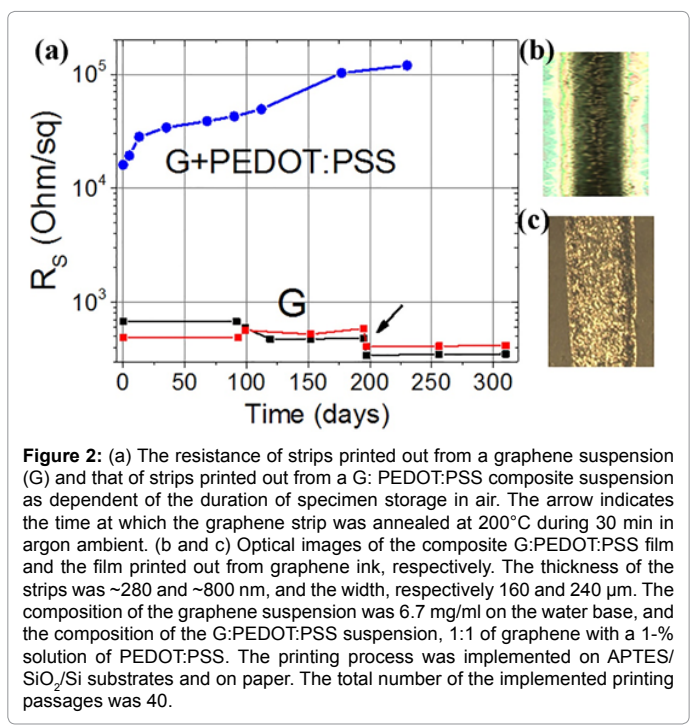

**Figure 2:** (a) The resistance of strips printed out from a graphene suspension (G) and that of strips printed out from a G: PEDOT:PSS composite suspension as dependent of the duration of specimen storage in air. The arrow indicates the time at which the graphene strip was annealed at 200°C during 30 min in argon ambient. (b and c) Optical images of the composite G:PEDOT:PSS film and the film printed out from graphene ink, respectively. The thickness of the strips was ~280 and ~800 nm, and the width, respectively 160 and 240 μm. The composition of the graphene suspension was 6.7 mg/ml on the water base, and the composition of the G:PEDOT:PSS suspension, 1:1 of graphene with a 1-% solution of PEDOT:PSS. The printing process was implemented on APTES/$SiO_2$/Si substrates and on paper. The total number of the implemented printing passages was 40.

### Fluorination of graphene suspensions

As it was shown in publication [16], the fluorination process begins at the edges of graphene flakes, at domain boundaries, and at defects. That is why the use of thin graphene flakes and fine graphene particles considerably enhances the fluorination process of graphene suspensions. In suspensions used immediately after the electrochemical exfoliation, flakes are comparatively large and, as a result, fluorographene will only be obtained after a treatment in an aqueous solution of hydrofluoric acid lasting for as long as 5-6 months. On the contrary, fine and thin graphene flakes can be fluorinated in 5-10 days. In the case of the suspension described in the previous section, the duration of the fluorination in an aqueous solution of hydrofluoric acid to a fluorination degree over 30% did not exceed 10 days. After the fluorination of the graphene suspension, the solvent was replaced with water to obtain from the fluorographene a water-based ink.

An HRTEM image of a dielectric layer printed out from a fluorinated graphene suspension is shown in Figure 3a. Crystallographic planes in the particles sized 3 to 5 nm (the particles are enclosed in circles) are distinctly seen. The interplanar spacing for the crystallographic planes normal to the image plane are somewhat larger than that for (002) crystallographic planes in graphite since, here, the Fourier spectrum reflections (Figure 3a) lie not in the circumference corresponding to the location of (002) reflections in graphite but closer to the center. A similar effect was also observed in the experimental electron microdiffraction patterns taken from these films (Figure 3d). The shift of the film-induced reflections inward the red circles for polycrystalline graphite (Figure 3d) shows that, as it could be expected, the FG lattice constant somewhat (by ~2%) exceeded the lattice constant in graphite. Thus, it can be concluded that the observed nanocrystals were few layer FG nanocrystals whose basic planes were oriented normally to the substrate. The AFM image of the surface of this film demonstrates the presence of larger particles with sizes up to 400 nm (Figure 3b). On the whole, the FG suspension transmits light (Figure 3c), the transmission value in the visible range amounting to 96-98% at a film thickness of 20 nm [17]. Here, the fluorination degree as estimated from the deconvolution of the C1s peak in the XPS spectrum or from the magnitude of the F1s/C1s peak intensity ratio amounts to 21-23%. The intensity of the main G (1580 cm$^{-1}$) and 2D (~2700 cm$^{-1}$) peaks in the Raman spectrum for FG is extremely weak, this spectrum being dominated by the 2D' (2940 cm$^{-1}$) peak. For comparison, Figure 3e shows the Raman spectrum of a thin film printed out from a graphene suspension with 1- to 3-monolayer thick flakes. Here, the 2D' peak is a dominating one as well. The key to explain the brightness of the 2D band and other overtone bands in the Raman spectra is based on the Kramers-Heisenberg-Dirac theory [18]. Heller has demonstrated that the transition sliding, or the sliding phonon production, explains the origin of the extremely bright overtone 2D band in graphene, and in the case of small-flake graphene suspensions an intense 2D' (or D+G) band peaking at ~2940 cm$^{-1}$ could be observed as well. The domination of the 2D' peak in the Raman spectra of FG could be due to the fact that most of the FG flakes had thicknesses of 1- to 5-monolayers.

### Functional layers prepared from weakly fluorinated graphene suspensions

HRTEM image of FG flakes with size about 6 nm from a weakly fluorinated graphene suspension (here, the fluorination degree as determined from x-ray photoelectron emission data was about 15%) are shown in Figure 4a. The inset shows an enlarged image of a separate flake, the characteristic feature of which is a darker center and a light perimeter. Since fluorination begins at particle edges, in the case of partial fluorination of graphene particles the central part of the flakes (dark center in Figure 4a) remains a non-fluorinated region. Figure 4b shows the current-voltage (I-V) characteristics of films with different fluorination degrees. Evidently, the I-V characteristics for fluorination degrees ~10-15% exhibit a negative differential resistance (NDR) behavior, very important for some applications, in particular for THz-range detectors [19]. Films with such characteristics were analyzed in detail [20]. It was shown that the emergence of NDR is related with the formation of a multi-barrier system in partially fluorinated graphene films. As it follows from Figure 4 and from the simulation data of ref. [21], the position of the peak observed in the I-V characteristics exhibits a shift with decreasing the size of graphene quantum dots (or with increasing the fluorination degree of graphene) in the FG matrix towards smaller voltages. A comparison between similar films obtained by fluorination of mechanically exfoliated graphene (in the case of few layer graphene, the quantum dots in different layers are vertically ordered) and films printed out from partially fluorinated graphene suspensions reveals the following regularity: the NDR behavior is only exhibited by the films obtained from the suspension. A distinguishing feature of the latter films is the overlap of graphene quantum dots in different layers. Very likely, in the cases of vertical ordering of graphene quantum dots the width of the FG-induced barriers was large in comparison with the thickness of the films in which the quantum dots in different layers were overlapping; the latter circumstance considerably decreases the effective width of the FG-induced potential barriers.

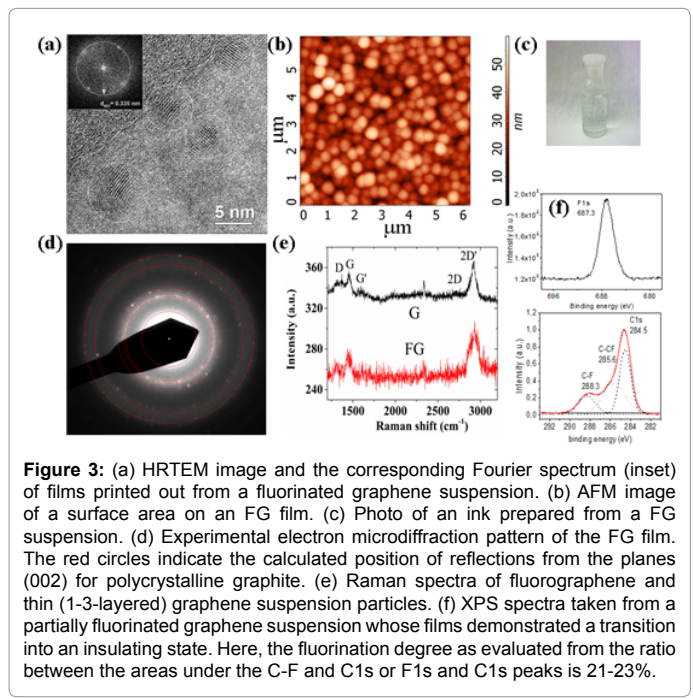

**Figure 3:** (a) HRTEM image and the corresponding Fourier spectrum (inset) of films printed out from a fluorinated graphene suspension. (b) AFM image of a surface area on an FG film. (c) Photo of an ink prepared from a FG suspension. (d) Experimental electron microdiffraction pattern of the FG film. The red circles indicate the calculated position of reflections from the planes (002) for polycrystalline graphite. (e) Raman spectra of fluorographene and thin (1-3-layered) graphene suspension particles. (f) XPS spectra taken from a partially fluorinated graphene suspension whose films demonstrated a transition into an insulating state. Here, the fluorination degree as evaluated from the ratio between the areas under the C-F and C1s or F1s and C1s peaks is 21-23%.

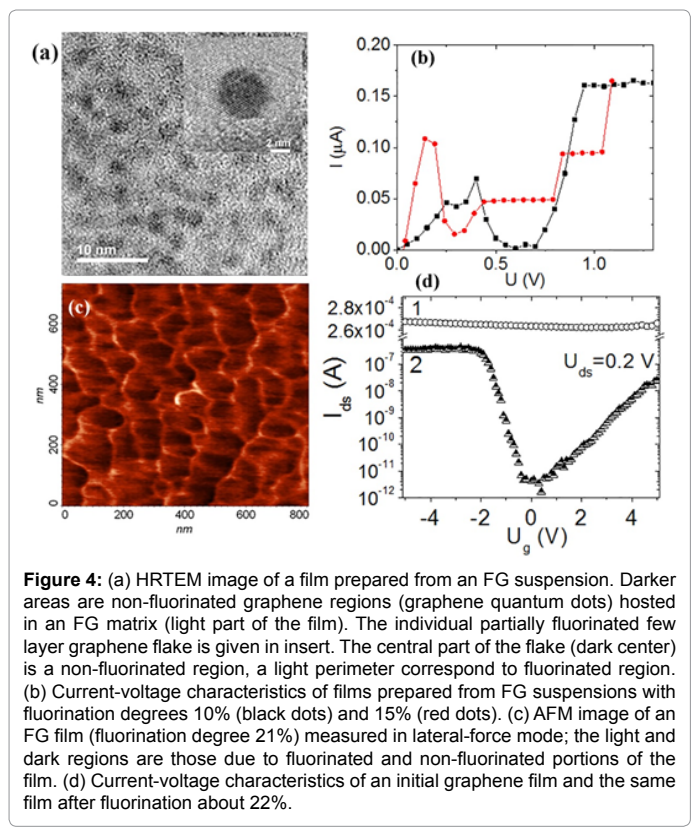

**Figure 4:** (a) HRTEM image of a film prepared from an FG suspension. Darker areas are non-fluorinated graphene regions (graphene quantum dots) hosted in an FG matrix (light part of the film). The individual partially fluorinated few layer graphene flake is given in insert. The central part of the flake (dark center) is a non-fluorinated region, a light perimeter correspond to fluorinated region. (b) Current-voltage characteristics of films prepared from FG suspensions with fluorination degrees 10% (black dots) and 15% (red dots). (c) AFM image of an FG film (fluorination degree 21%) measured in lateral-force mode; the light and dark regions are those due to fluorinated and non-fluorinated portions of the film. (d) Current-voltage characteristics of an initial graphene film and the same film after fluorination about 22%.

Another property important for applications of partially fluorinated graphene layers is observed in the cases of the fluorination degree about 21-23%. Here, the conduction in the material is still not blocked yet, but the FG related barriers have already partially formed. As a result, the transfer characteristics of transistor-like structures with partially fluorinated graphene channel exhibit a gate-voltage-induced current modulation reaching six orders of (Figures 4c and 4d). This effect is related with the passage of electric current across the tunnel-transparent barriers whose height was varied by the gate voltage. The latter phenomenon was observed in films that were mechanically exfoliated from graphite and subsequently given a fluorination treatment.

### Properties of printed-out FG dielectric layers and G/FG hetero-structures

From an FG suspension with fluorination degree 40-50%, metal-insulator-semiconductor (MIS) structures similar to the structures shown in Figure 5a were printed out. The thickness of the printed films was 20-40 nm. Capacitance-voltage measurements performed on such structures have shown that the dielectric films printed out from FG suspensions demonstrated an ultra-low built-in charge of $(0.5-4) \times 10^{10}$ $cm^{-2}$ and a roughly identical density of states at the hetero-boundary with semiconductors [11]. In those cases in which the dielectric films in MIS structures were obtained by printing, the magnitude of the built-in charges approached the lower boundary of the indicated range. In those cases in which dielectric films were deposited from droplets, the magnitude of the built-in charges was close to the upper boundary of the above range. It should be noted that, as a rule, advanced thin (10-40 nm thick) dielectric films ($Al_2O_3$, $HfO_2$, $SiO_2$, graphene oxide, etc.) exhibit much higher (by 1.5-2 orders) charge values. Thus, the dielectric films prepared from fluoro-graphene possess ultra-low built-in charges.

Figures 6a-6c show Ag/FG/Ag structures printed out from silver and an FG ink. A schematic representation of those structures is shown in Figure 6a. In the SEM data of Figure 6e, one can see the edge of the thickest FG film (printed out in 20 printing passages). According to AFM data, the thickness of this film was 35 nm. Electrical measurements have shown that, in all the three types of printed Ag/FG/Ag structures with FG-layer thickness 20-40 nm, the leakage currents were smaller than $10^{-8}$ $A/cm^2$ at the breakdown field being higher than $10^8$ V/cm. The excellent combination of such properties, that is, an ultra-low magnitude of built-in charges, ultra-small leakage currents, and a very high breakdown field value makes the FG films promising ones as carrier layers for graphene in interface engineering, as the gate dielectric for transistor structures, etc. No other analogues to such films are presently available.

Fluorographene films obtained by 2D printing technology on solid and flexible substrates from such a suspension are shown in Figures 7a and 7b. The film thickness was (Figure 7a); a surface relief with a height within 2.2 nm was observed. According to AFM data, the roughness of the film surface in the image of Figure 7a was 0.3 nm; in this case, the characteristic particle sizes were smaller than 100 nm at particle thicknesses falling in the range from one monolayer to 2 nm. The decrease of suspension-particle sizes was due to the additional splitting and fragmentation of the particles stimulated by non-uniform mechanical stresses induced by the fluorination process.

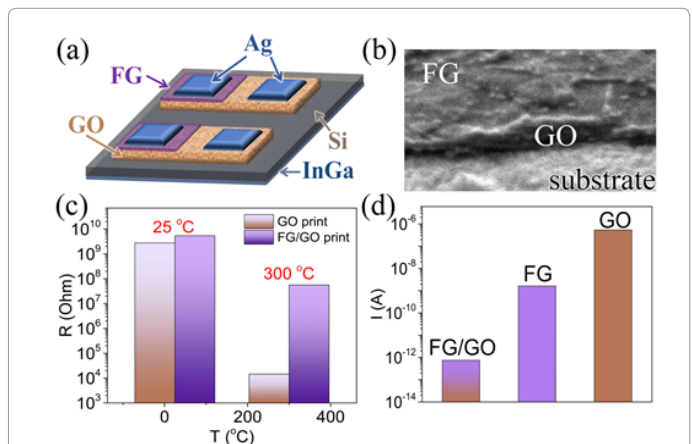

**Figure 5:** (a) Test MIS structures with gate dielectrics formed from FG/GO and GO films. (b) SEM image of the two-layer film. (c) Histograms of resistance values in the MIS structures with the FG/GO and GO films before and after an annealing performed at 300°C during 30 min in argon ambient. (d) Histograms of leakage currents across the FG/GO bilayer films, and histograms of leakage currents across the component GO and FG films in the structure.

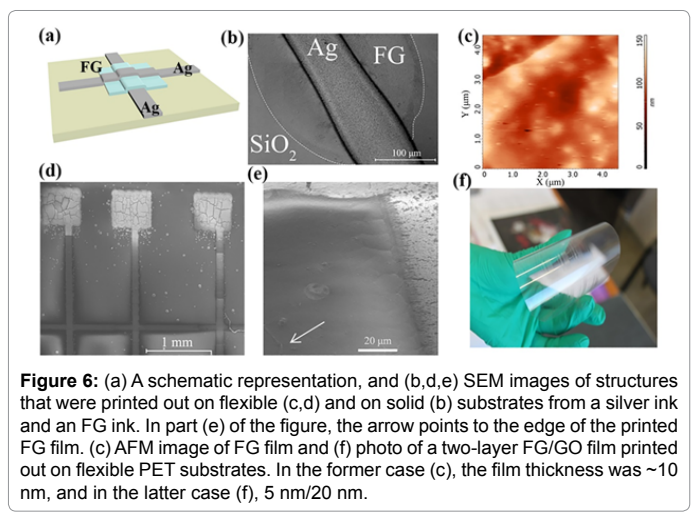

**Figure 6:** (a) A schematic representation, and (b,d,e) SEM images of structures that were printed out on flexible (c,d) and on solid (b) substrates from a silver ink and an FG ink. In part (e) of the figure, the arrow points to the edge of the printed FG film. (c) AFM image of FG film and (f) photo of a two-layer FG/GO film printed out on flexible PET substrates. In the former case (c), the film thickness was ~10 nm, and in the latter case (f), 5 nm/20 nm.

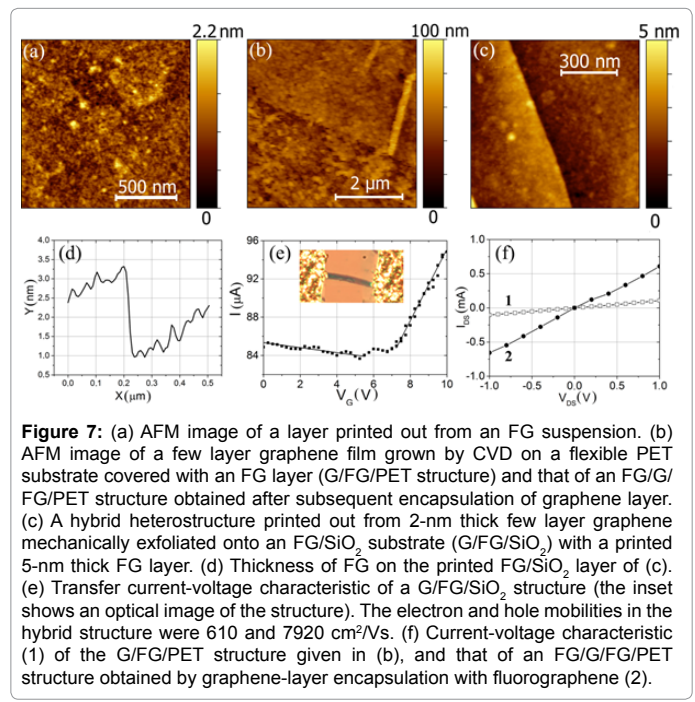

**Figure 7:** (a) AFM image of a layer printed out from an FG suspension. (b) AFM image of a few layer graphene film grown by CVD on a flexible PET substrate covered with an FG layer (G/FG/PET structure) and that of an FG/G/FG/PET structure obtained after subsequent encapsulation of graphene layer. (c) A hybrid heterostructure printed out from 2-nm thick few layer graphene mechanically exfoliated onto an FG/SiO$_2$ substrate (G/FG/SiO$_2$) with a printed 5-nm thick FG layer. (d) Thickness of FG on the printed FG/SiO$_2$ layer of (c). (e) Transfer current-voltage characteristic of a G/FG/SiO$_2$ structure (the inset shows an optical image of the structure). The electron and hole mobilities in the hybrid structure were 610 and 7920 cm$^2$/Vs. (f) Current-voltage characteristic (1) of the G/FG/PET structure given in (b), and that of an FG/G/FG/PET structure obtained by graphene-layer encapsulation with fluorographene (2).

Then, mechanically exfoliated few layer graphene of thickness 2 nm was applied onto this FG film on a SiO$_2$/Si substrate (Figures 7c and 7d). As a result, a hybrid heterostructure few layer graphene on FG/SiO$_2$/Si substrate was obtained. The transfer characteristic of such a G/FG/SiO$_2$/Si transistor-like structure obtained using the silicon substrate as the gate is shown in Figure 7e. The mobility of charge carriers for such a structure was 600 and 8000 cm$^2$/Vs for electrons and holes, respectively. In similar structures created without an FG sublayer the mobility was 800-1200 cm$^2$/Vs. The comparison is therefore indicative of a substantial increase of electron mobility in the structures containing an FG printed sublayer.

In the case of the flexible G/FG/PET heterostructure (Figure 7a), the encapsulation of the CVD-grown graphene film with an FG layer (FG/G/FG/PET structure, Figures 7b and 7f) has led to an increase of graphene conductivity by 5-6 times. On the whole, the obtained data show that the combined use of fluorographene with the graphene of hybrid heterostructures offers promise in applications.

Now we consider bilayer structures with 3-5 nm thick FG layer printed out on 25-nm thick graphene oxide films (Figure 5). The process yielding graphene oxide layers, the so-called modified Hammers method, is a well-established technique; yet, using graphene oxide as dielectric layers is hampered by instability of GO properties and by large leakage currents [22]. Nonetheless, attempts were made toward using graphene oxide as the gate dielectric in flexible transistor structures since other dielectrics, especially boron nitride, cannot withstand expansion or bending [23,24]. That is why we have tried to remedy those deficiencies of GO. That was made by creating the FG/GO bilayer films and MIS structures. Having made a comparison between the resistances of the GO and FG/GO films, we have found that the resistance had somewhat increased in value after the application of the thin film (Figure 5c). After an annealing of GO films at temperature 300 °C during 30 min in argon ambient, the resistance of the GO film has decreased by more than five orders because of the partial reduction of the graphene oxide. On the other hand, the resistance of the bilayer film had only shown a change by one and a half order, thus indicating that a thin FG film impedes the out-diffusion of oxygen and the reduction of GO. Figure 5d shows histograms that illustrate the change of leakage current in the bilayer film in comparison with the one-layer GO or FG films. Evidently, on application of an FG film as thin as several nanometers the leakage current across the composite film has decreased almost by six orders. It should be noted here that, normally, the leakage currents across GO are nearly independent of the film thickness, at least in the range of thicknesses 20 to 100 nm [25]. Supposedly, this phenomenon was related with the decoration of structural defects in GO with finer FG particles. On the whole, it is evident that application of a FG layer onto the film surface stabilized the GO film. Other properties of the bilayer films were discussed in more detail [17].

### Composite films exhibiting a resistive switching behavior

Memory devices functioning on the resistive switching phenomenon attracts ever-increasing attention due to their simplicity,

low energy consumption, low writing voltages, short switching times, and the possibility of non-volatile, long-term data storage [26,27]. By now, only few materials exhibit a resistive switching behavior can be used for fabrication of memory cells in the field of flexible and printed electronics. Graphene oxide is a material that is used most frequently [28,29]; however, under the action of electric field and temperature this material looses oxygen-containing functional groups and, for this reason, graphene-oxide-based devices will most likely turn to be unstable. Besides, as it was noted above, graphene oxide layers normally exhibit large leakage currents. As an alternative, we have attempted using fluorinated graphene as a more stable graphene derivative.

Figure 8 show current-voltage characteristics of two types of structures. In the first case (Figure 8a), the structures were created by printing of a thin FG film on the surface of polyvinyl alcohol film. The fluorination degree of the fluorographene was ~20%. The obtained bilayer FG/PVA film was located on a silicon substrate, and the lower PVA layer was applied using a spin-processor. On the surface of the FG/PVA film, silver contacts were prepared. The second contact was connected to the substrate, so that a MIS structure was formed. The current-voltage characteristics exhibited a stable phenomenon of unipolar resistive switching whose magnitude was up to one order. Repeatedly performed measurements (up to 100 measurements with subsequent measurements performed over a period of half a year) have proved both stability of such structures and stability of the exhibited resistive switches.

In the latter case (Figure 8b); the films were prepared by printing crossbar structures (Figure 5a) from an FG:$V_2O_5$ composite suspension on both solid and flexible substrates. The suspension contained fluorinated graphene and vanadium oxide (predominantly, $V_2O_5$) nanoparticles. The lower and upper contacts were printed out from an ink that contained silver particles. The current-voltage characteristics of the crossbar structures exhibited a stable resistive switching behavior with current modulation reaching five orders. According to SEM and ACM data, the sizes of $V_2O_5$ particles were smaller than 50 nm. On mixing with the FG suspension, the encapsulation of $V_2O_5$ particles with fluorographene was assumed.

Studies to reveal the nature of the resistive switching phenomenon in two-layer and composite FG-based films are presently under way. However, by now considerable potential of such functional layers for the development of resistive-memory materials is quite evident, including the case of flexible substrates.

## Conclusion

Using the 2D printing technology, heterostructures containing graphene and fluorographene layers printed out on solid and flexible substrates were fabricated. Considerable potential of such heterostructures in electronic applications was demonstrated. In stability studies of printed graphene layers about 20 nm thick, it was found that the electrical resistance of such layers showed no substantial changes over a period of one year. In the case of composite graphene:PEDOT:PSS layers, their electrical resistance has increased by a factor of 2.5 during four month of specimen storage in air ambient. Unique properties of fluorographene (at fluorination degree above 30%) as an insulating layer for graphene-based heterostructures were revealed. It was shown that the leakage currents in thin (20-40 nm) FG films obtained by the 2D printing method could be made lower than $10^{-8}$ A/cm$^2$, whereas the breakdown field exceeded $10^8$ V/cm. The latter properties of the FG films are combined with their ultra-low built-in charges, which combination makes the films suitable for application as the gate dielectric in FET structures. In hybrid structures with printed FG layers in which graphene was transferred onto, or encapsulated with, an FG layer, an increase of charge-carrier mobility and material conductivity in the graphene layer amounting to 5-6 times was observed. The spectrum of future applications of FG layers extends due to the possibility of obtains from suspensions weakly fluorinated graphene functional layers. Fluorinated graphene fluorination degree <20% exhibit a negative differential resistance behavior. At fluorination degrees of 20-23%, FET channels with a current modulation amounting to 5-6 orders and resulting from the formation of a multi-barrier system in which the electric current is defined by the height of the built-in potential barriers and, hence, can be modulated by the gate voltage. Composite or two-layer films based on fluorographene and $V_2O_5$ particles suspended in polyvinyl alcohol exhibit a resistive switching behavior. For various composites, the resistance variation amounts to 1 to 5 orders. On the whole, the investigated graphene/FG heterostructures show considerable potential in many applications, including flexible electronics.

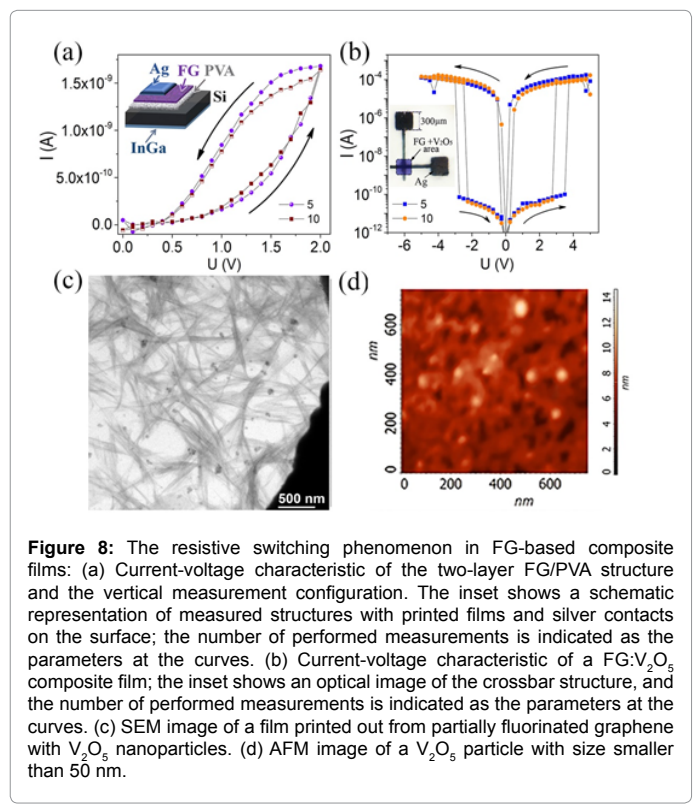

**Figure 8:** The resistive switching phenomenon in FG-based composite films: (a) Current-voltage characteristic of the two-layer FG/PVA structure and the vertical measurement configuration. The inset shows a schematic representation of measured structures with printed films and silver contacts on the surface; the number of performed measurements is indicated as the parameters at the curves. (b) Current-voltage characteristic of a FG:$V_2O_5$ composite film; the inset shows an optical image of the crossbar structure, and the number of performed measurements is indicated as the parameters at the curves. (c) SEM image of a film printed out from partially fluorinated graphene with $V_2O_5$ nanoparticles. (d) AFM image of a $V_2O_5$ particle with size smaller than 50 nm.

## Acknowledgements

The authors express their gratitude to Dr. S.A. Smagulova, Scientific Researcher of the Ammosov North-Eastern Federal University, for useful assistance and preparation of oxidized graphene suspensions and to Dr. V.B. Timofeev, from the same University, for his kind assistance in performing SEM measurements. This study was financially supported in part by the Russian Science Foundation (Grant No. 15-12-00008). HRTEM studies were performed using the equipment of CCU "Nanostructures" with the support of RSCF (project No 14-22-00143).

## References

1. Zhao J, Zhang GY, Shi DX (2013) Review of graphene-based strain sensors. Chin Phys B 22: 057701.

2. Nair RR, Ren W, Jalil R, Riaz I, Kravets VG, et al. (2010) Fluorographene: a two dimensional counterpart of Teflon. Small 6: 2877-2884.

3. McManus D, Vranic S, Withers F, Sanchez-Romaguera V, Macucci M, et al. (2017) Water-based and biocompatible 2D crystal inks for all-inkjet-printed heterostructures. Nat Nanotech 12: 343-350.

4. Kamyshny A, Magdassi S (2014) Conductive nanomaterials for printed electronics. Small 10: 3515-3535.

5. Magliulo M, Mulla MY, Singh M, Macchia E, Tiwari A, et al. (2015) Printable and flexible electronics: from TFTs to bioelectronic devices. J Mater Chem C 3: 12347-12363.

6. Kim J, Son D, Lee M, Song C, Song J-K, et al. (2016) A wearable multiplexed silicon nonvolatile memory array using nanocrystal charge confinement. Sci Advances 2.

7. Antonova IV (2017) 2D printing technologies using graphene based materials. Physics Uspechi 60: 204-218.

8. Wu W (2017) Inorganic nanomaterials for printed electronics: a review. Nanoscale 9: 7342-7372.

9. Nebogatikova NA, Antonova IV, Volodin VA, Prinz VY (2013) Functionalization of graphene and few-layer graphene with aqueous solution of hydrofluoric acid. Phys E 52: 106-111.

10. Nebogatikova NA, Antonova IV, Prinz VY, Kurkina II, Alexandrov GN, et al. (2015) Fluorinated graphene dielectric films obtained from functionalized graphene suspension: preparation and properties. Phys Chem Chem Phys 17: 13257-13266.

11. Antonova IV, Nebogatikova NA (2017) Fluorinated Graphene Dielectric and Functional Layers for Electronic Applications. In Graphene Materials-Advanced Applications. InTech pp. 211-230.

12. Withers F, Bointon TH, Dubois M, Russo S, Craciun MF (2011) Nanopatterning of Fluorinated Graphene by Electron Beam Irradiation. Nano Lett 11: 3912-3916.

13. Yakimchuk EA, Soots RA, Antonova IV (2017) 2D printed graphene conductive layers with high carrier mobility. Current Appl Phys.

14. Mengistie DA, Ibrahem MA, Wang PC, Chu CW (2014) Highly Conductive PEDOT:PSS Treated with Formic Acid for ITO-Free Polymer Solar Cells. ACS Appl Mater Interfaces 6: 2292-2299.

15. Liu Z, Parvez K, Li R, Dong R, Feng X, et al. (2014) Transparent Conductive Electrodes from Graphene/PEDOT:PSS Hybrid Inks for Ultrathin Organic Photodetectors. Adv Mater 27: 669-675.

16. Nebogatikova NA, Antonova IV, Prinz VY, Timofeev VB, Smagulova SA (2014) Graphene quantum dots in fluorographene matrix formed by means of chemical functionalization. Carbon 77: 1095-1103.

17. Ivanov AI, Nebogatikova NA, Kotin IA, Antonova IV (2017) Two-layer and composite films based on oxidized and fluorinated graphene. Phys Chem Chem Phys 19: 19010-19020.

18. Heller EJ, Yang Y, Kocia L, Chen W, Fang S, et al. (2016) Theory of Graphene Raman Scattering. ACS Nano 10: 2803-2818.

19. Ryzhii V, Dubinov AA, Otsuji T, Aleshkin VY, Ryzhii M, et al. (2013) Double-graphene-layer terahertz laser: concept, characteristics, and comparison. Opt Exp 21: 31567-31577.

20. Antonova IV, Kurkina IV, Nebogatikova NA, Komonov AI, Smagulova SA (2017) Films fabricated from partially fluorinated graphene suspension: structural, electronic properties and negative differential resistance. Nanotechnology 27: 074001.

21. Antonova IV, Shojaei S, Sattari-Esfahlan SM, Kurkina (2017) Negative differential resistance in partially fluorinated graphene films. Appl Phys Let 111: 043108.

22. Lee SK, Jang HY, Jang S, Choi E, Hong BH, et al. (2012) All graphene-based thin film transistors on flexible plastic substrates. Nano Lett 12: 3472-3476.

23. Petrone N, Chari T, Meric I, Wang L, Shepard KL, et al. (2015) Flexible graphene field-effect transistors encapsulated in hexagonal boron nitride. ACS Nano 9: 8953-8959.

24. Petrone N, Meric I, Hone J, Shepard KL (2013) Graphene field-effect transistors with gigahertz frequency power gain on flexible substrates. Nano Lett 13: 121-125.

25. Antonova IV (2016) Non-Organic Dielectric Layers for Graphene and Flexible Electronics. International Journal of Nanomaterials. Nanotechnology and Nanomedicine 2: 18-24.

26. Yang JJ, Strukov DB, Stewart DR (2013) Memristive devices for computing. Nat Nanotech 8: 13-24.

27. Worfolk BJ, Andrews SC, Park S, Reinspach J, Liu N, et al. (2015) Ultrahigh electrical conductivity in solution-sheared polymeric transparent films PNAS. Proceedings of the National Academy of Sciences 112: 14138-14143.

28. Jeong HY, Kim JY, Kim JW, Hwang JO, Kim JE, et al. (2010) Graphene oxide thin films for flexible nonvolatile memory applications. Nano Lett 10: 4381-4386.

29. Lin CC, Wu HY, Lin NC, Lin CH (2014) Graphene-oxide-based resistive switching device for flexible nonvolatile memory application. Jpn J App Phys 53.